\documentclass[11pt]{article}
\usepackage{graphicx} 	
\usepackage{amsmath} 	
\usepackage{array} 	
\usepackage{hhline} 	
\usepackage{times}      
\usepackage{amsthm} 
\usepackage{verbatim}
\usepackage{fullpage}
\usepackage{url}
\def\MyStretch{1.1}
\def\baselinestretch{\MyStretch}

\newcommand{\urlBiBTeX}[1]{\url{#1}} 

\newtheorem{theorem}{Theorem}
\newtheorem{lemma}{Lemma}
\def\Pr{\mbox{Pr}}

\def\mynormality#1{\def\baselinestretch{#1}\small\normalsize}
\mynormality{1.0}

\author{
{\em Nicolas Christin} \hspace{.5cm} {\em John Chuang}\\
School of Information Management and Systems\\
University of California, Berkeley\\
Berkeley, CA 94720\\
{\tt\{christin,chuang\}@sims.berkeley.edu}
}

\title{\bf On the cost of participating in a peer-to-peer 
network
\thanks{This research is supported in part by the
National Science Foundation, through grant ANI-0085879.}}

\begin{document}
\date{
{\small\em Technical report, University of California, Berkeley\\\rm\url{http://p2pecon.berkeley.edu/pub/TR-2003-12-CC.pdf}}
\\\vspace{.5cm}
December 2003}
\maketitle
\thispagestyle{empty}

\begin{abstract}
In this paper, we model the cost incurred by each peer
participating in a peer-to-peer network. Such a cost model allows to
gauge potential disincentives for peers to collaborate, and provides
a measure of the ``total cost'' of a network, which is a possible
benchmark to distinguish between proposals. We characterize the cost
imposed on a node as a function of the experienced load and the node
connectivity, and show how our model applies to a few proposed routing
geometries for distributed hash tables (DHTs). We further outline a
number of open questions this research has raised.
\end{abstract}
\mynormality{1.2}
\newpage
\section{Introduction}
\label{sec:intro}
A key factor in the efficiency of a peer-to-peer overlay network is the
level of collaboration provided by each peer. This paper takes a first step
towards quantifying the level of collaboration that can be expected from
each participant, by proposing a model to evaluate the cost each peer
incurs for being a part of the overlay.

Such a cost model has several useful applications, among which,
(1) providing a benchmark that can be used to compare between different
proposals, complementary to recent works comparing topological properties
of various overlays \cite{Gummadi:SIGCOMM03, Loguinov:SIGCOMM03}, (2)
allowing for predicting disincentives, and designing mechanisms that
ensure a protocol is {\em strategyproof} \cite{Ng:p2pecon03}, 
and (3) facilitating the design of load balancing primitives.

This work is not the first attempt to characterize the cost of
participating in a network. Jackson and Wolinsky \cite{JackWol96}
proposed cost models to analyze formation strategies in social
and economic networks. More recent studies \cite{Chun:INFOCOM04,
Fabrikant:PODC03} model (overlay) network formation as a non-cooperative
game. These studies assume that each node has the freedom to choose
which links it maintains, whereas we assume that the overlay topology
is constrained by a protocol. Moreover, our approach extends previously
proposed cost models \cite{Chun:INFOCOM04, Fabrikant:PODC03, JackWol96},
by considering the load imposed on each node in addition to the distance
to other nodes and degree of connectivity.

In the remainder of this paper, we introduce our proposed cost model,
before applying it to several routing geometries used in recently
proposed distributed hash tables (DHT) algorithms \cite{Koorde,
Loguinov:SIGCOMM03, CAN, Pastry, Chord:TON}. We conclude by discussing
some open problems this research has uncovered.
\section{Proposed cost model}
\label{sec:cost-model}
The model we propose applies to any peer-to-peer network where nodes
request and serve items, or serve requests between other nodes. This
includes peer-to-peer file-sharing systems \cite{gnutella}, ad-hoc
networks \cite{adhoc}, peer-to-peer lookup services \cite{CAN,
Chord:TON}, peer-to-peer streaming systems \cite{PROMISE}, or
application-layer multicast overlays \cite{NICE:SIGCOMM02, Narada,
LiNaSi02}, to name a few examples.

To simplify the presentation, we assume a DHT-like structure, defined
by quadruplet $(V, E, K, F)$, where $V$ is the set of vertices in the
network, $E$ is the set of edges, $K$ is the set of keys (items) in the
network, and $F: K \rightarrow V$ is the hash function that assigns
keys to vertices. We denote by $K_i = \{k \in K : F(k) = i \}$ the
set of keys stored at node~$i \in V$. We have $K = \bigcup_{i} K_i$,
and we assume, without loss of generality, that the sets $K_i$ are
disjoint.\footnote{If a key is stored on several nodes (replication),
the replicas can be considered as different keys with the exact same
probability of being requested.} We characterize each request with two
independent random variables, $X \in V$ and $Y \in K$, which denote
the node~$X$ making the request, and the key~$Y$ being requested,
respectively.

Consider a given node $i \in V$. Every time a key $k$ is requested in
the entire network, node~$i$ is in one of four situations:
\begin{enumerate}
\item{
Node~$i$ does not hold or request $k$, and is not on the routing path
of the request. Node~$i$ is not subject to any cost.
}
\item{
Node~$i$ holds key $k$, and pays a price $s_{i,k}$ for
serving the request. We define the {\em service cost} $S_i$ incurred
by~$i$, as the expected value of $s_{i,k}$ over all possible requests.
That is,
$$
S_i = \sum_{k \in K_i}s_{i,k}\Pr[Y = k] \ . 
$$
}
\item{
Node~$i$ requests key $k$, and pays a price to look up and retrieve~$k$. We
model this price as $a_{i,k}t_{i,j}$, where $t_{i,j}$ is the number of
hops between $i$ and the node~$j$ that holds the key~$k$, and $a_{i,k}$
is a (positive) proportional factor. We define the {\em access cost} 
suffered by
node~$i$, $A_i$, as the sum of the individual costs $a_{i,k}t_{i,j}$
multiplied by the probability key~$k \in K_j$ is requested, that is,
\begin{equation}
A_i = \sum_{j \in V}\sum_{k\in K_j} a_{i,k}t_{i,j}  \Pr [Y=k] \ ,
\label{eq:access}
\end{equation}
with $t_{i,j} = \infty$ if there is no path from node~$i$ to node~$j$,
and $t_{i,i} = 0$ for any~$i$.
}
\item{
$i$ does not hold or request $k$, but has to forward the request for
$k$, thereby paying a price $r_{i,k}$. The overall {\em routing cost}
$R_i$ experienced by node~$i$ is the average over all possible keys~$k$,
of the values of 
$r_{i,k}$ such that $i$ is on the path of the request. That is,
we consider the binary function
$$
\chi_{j,l}(i) = \left\{
\begin{array}{ll}
1 & \mbox{if $i$ is on the path from $j$ to $l$,}\\
& \mbox{excluding $j$ and $l$}\\
0 & \mbox{otherwise,}
\end{array}
\right .
$$
and express $R_i$ as
\begin{equation}
R_i = \sum_{j\in V}\sum_{l\in V}\sum_{k\in K_l}r_{i,k}\Pr [X = j]  \Pr[Y=k] \chi_{j,l}(i)\ .
\label{eq:routing}
\end{equation}
}
\end{enumerate}
In addition, each node keeps some state information so that the protocol
governing the DHT operates correctly. In most DHT algorithms, each
node~$i$ maintains a neighborhood table, which grows linearly with
the out-degree $\deg(i)$ of the node, resulting in a {\em maintenance
cost}~$M_i$ given by
$$
M_i = m_i \deg(i) \ ,
$$
where $m_i > 0$ denotes the cost of keeping a single entry in the
neighborhood table of node~$i$.

Last, the {\em total cost} $C_i$ imposed on node~$i$ is given by 
$$
C_i = S_i+A_i+R_i+M_i \ ,
$$
which can be used to compute the total cost of the network, $C = \sum_{i
\in V} C_i$. The topology that minimizes $C$, or ``social optimum,'' 
is generally not trivial. In particular, the social optimum is the
full mesh only if $m_i = 0$ for all~$i$, and the empty set only if
$a_{i,k}=0$ for all~$(i,k)$.
\section{Case studies}
\label{sec:case-studies}
We next apply the proposed cost model to a few selected routing
geometries. We define a routing geometry as in \cite{Gummadi:SIGCOMM03},
that is, as a collection of edges, or topology, associated with a
route selection mechanism. Unless otherwise noted, we assume shortest
path routing, and distinguish between different topologies. We derive
the various costs experienced by a node in each geometry, before
illustrating the results with numerical examples.

\subsection{Analysis}
\label{subsec:formal}
We consider a network of $N > 0$ nodes, and, for simplicity, assume
that, for all $i$ and $k$, $s_{i,k} = s$, $a_{i,k} = a$, $r_{i,k} = r$,
and $m_i = m$. For the analysis in this section, we also assume that
each node holds the same number of keys, and that all keys have the same
popularity. As a result, for all $i$, 
$$
\sum_{k\in K_i} \Pr [Y=k] = \frac{1}{N} \ , 
$$ 
which implies 
$$
S_i = \frac{s}{N} \ ,
$$ 
regardless of the geometry considered. We also
assume that requests are uniformly distributed over the set of nodes, that is, 
for any node~$i$, 
$$\Pr [X=i] = \frac{1}{N} \ . 
$$ 
Last, we assume that
no node is acting maliciously.

\paragraph{Star network}
The star frequently appears 
as an equilibrium in network formation studies using cost models based on
graph connectivity \cite{Chun:INFOCOM04, Fabrikant:PODC03, JackWol96}.

We use $i = 0$ to denote the center of the star,  
which routes all traffic between peripheral nodes. That is,
$\chi_{j,l}(0) = 1$ for any $j \neq l$ ($j>0$, $l > 0$). 
Substituting
in Eqn.~(\ref{eq:routing}), we get
$$
R_0 = \frac{r(N-1)(N-2)}{N^2} \ . 
$$
The center node is located at a distance of one hop from all $(N-1)$
other nodes, thus
$$
A_0 = \frac{a(N-1)}{N} \ .
$$ 
In addition, $\deg(0) = N-1$, which implies that 
the cost incurred by the center of the star, $C_0$, is
\begin{equation}
C_0 = m(N-1)+\frac{s}{N}+\frac{a(N-1)}{N}+\frac{r(N-1)(N-2)}{N^2} \ .
\label{eq:c0-star}
\end{equation}

Peripheral nodes do not route any traffic, i.e., $R_i = 0$ for all
$i>0$, and are located at a distance of one from the center of the star,
and at a distance of two from the $(N-2)$ other nodes, giving 
$$A_i = \frac{a(2N-3)}{N} \ .
$$ 
Furthermore, $\deg(i) = 1$ for all peripheral nodes. Thus, $M_i = m$, and 
the total cost imposed on nodes $i > 0$ is 
\begin{equation}
C_i = m+\frac{s+a(2N-3)}{N} \ .
\label{eq:ci-star}
\end{equation}
The difference $C_0 - C_i$ quantifies the (dis)incentive to be in the
center of the star. As expressed in the following two theorems, 
there is a (dis)incentive to be in the center of
the star in a vast majority of
cases.
\begin{theorem}
If the number of nodes $N$ ($N>0$) is variable, $C_0 \neq C_i$ 
unless $m = r = a = 0$.
\label{theo:star-asymmetry-x1}
\end{theorem}
\begin{proof}
Assume that $C_0 - C_i = 0$. Because $N \neq 0$, 
$C_0 - C_i = 0$ is equivalent to $N^2(C_0-C_i) = 0$.
Using the expressions
for $C_0$ and $C_i$ given in Eqs.~(\ref{eq:c0-star}) and
(\ref{eq:ci-star}), and rewriting the condition $N^2(C_0-C_i) = 0$
as a polynomial in $N$, we obtain
$$
mN^3-(2m+a-r)N^2+(2a-3r)N+2r = 0 \ .
$$
We can factor the above by
$(N-2)$, and obtain
\begin{equation}
(N-2)(mN^2-(a-r)N-r) =  0 \ .
\label{eq:polynomial}
\end{equation}
A polynomial in $N$ is constantly equal to zero if and only if
all of the polynomial coefficients are equal to zero. Thus, 
Eqn.~(\ref{eq:polynomial}) holds for {\em any} value of $N$ if and only if:
$$
\left\{
\begin{array}{lll}
m &=& 0 \ ,\\
a-r &=& 0 \ ,\\
r &=& 0 \ .\\
\end{array}
\right .
$$
The solutions of the above system of equations are $m = r = a =
0$. Hence, $C_0 - C_i = 0$ for any $N$ only when nodes only pay an
(arbitrary) price for serving data, while state maintenance, traffic
forwarding, and key lookup and retrieval come for free.
\end{proof}
\begin{theorem}
If the number of nodes $N$ ($N > 0$) is held fixed, and at least one of $m$, $r$, or $a$ is different from zero, $C_0 = C_i$ only if 
$N = 2$ or $N = N_0$, where $N_0$ is a
positive integer that must satisfy:
\begin{equation}
N_0 = \left\{
\begin{array}{ll}
\frac{r}{r-a} 
&\mbox{if $m=0$ and $r \neq a$} \ ,\\
\frac{a-r}{2m} + \sqrt{\left(\frac{a-r}{2m}\right)^2+\frac{r}{m}} 
&\mbox{if $m\neq 0$} \ .
\end{array}
\right .
\label{eq:n0}
\end{equation} 
Additionally, $C_0 \neq C_i$ for any $N \neq 2$ if $m=0$ and $r=a$.
\label{theo:star-asymmetry-x2}
\end{theorem}
\begin{proof}
Recall from the proof of Theorem~\ref{theo:star-asymmetry-x1}, that 
$C_0-C_i = 0$ is equivalent to Eqn.~(\ref{eq:polynomial}).
Clearly, setting $N=2$ satisfies Eqn.~(\ref{eq:polynomial}) for all
values of $s$, $r$, and $m$.
Assuming now that $N\neq 2$, to have $C_0-C_i = 0$, we need to have 
\begin{equation}
mN^2-(a-r)N-r = 0 \ .
\label{eq:degree2}
\end{equation}
Since at least one of $m$, $r$, or $a$ is not equal to zero, 
Eqn.~(\ref{eq:degree2}) 
has at most two real solutions. We distinguish between all possible
cases for $m$, $r$, and $a$ such that at least one of $m$, $r$, and $a$
is different from zero.
\begin{itemize}
\item{If $m = 0$, and $a=r$, Eqn.~(\ref{eq:degree2}) reduces to $r =
0$, which implies $m = r = a = 0$, thereby contradicting the hypothesis
that at least one of $m$, $r$, and $a$ is different from zero. Therefore,
Eqn.~(\ref{eq:degree2}) does not admit any solution, i.e., there is
a (dis)incentive to be in the center of the star regardless of~$N$.}
\item{If $m = 0$ and $r \neq a$, the only solution to        
Eqn.~(\ref{eq:degree2}) is 
\begin{equation}
N_0 = \frac{r}{r-a} \ . 
\label{eq:degree1-sol}
\end{equation}
Note that if $r <     
a$, $N_0 < 0$ which is not feasible. (The number of nodes has to be     
positive.)}
\item{If $m \neq 0$, then Eqn.~(\ref{eq:degree2}) admits two real roots
(or a double root if $a=r=0$), given by 
$$
N_0 = 
\frac{a-r}{2m} \pm \sqrt{\left(\frac{a-r}{2m}\right)^2+\frac{r}{m}}\ . 
$$
However, because $r \geq 0$, and $m\geq 0$, 
$$
\frac{a-r}{2m} -
\sqrt{\left(\frac{a-r}{2m}\right)^2+\frac{r}{m}} \leq 0 \ , 
$$
so that the
only potentially feasible $N_0$ is given by 
\begin{equation}
N_0 = 
\frac{a-r}{2m} + \sqrt{\left(\frac{a-r}{2m}\right)^2+\frac{r}{m}}\ . 
\label{eq:degree2-sol}
\end{equation}
}
\end{itemize}
Combining Eqs.~(\ref{eq:degree1-sol}) and Eqs.~(\ref{eq:degree2-sol}) yields the expression for $N_0$ given in Eqn.~(\ref{eq:n0}). Note that the expression given in Eqn.~(\ref{eq:n0}) is only
a necessary condition. In addition, $N_0$ has to be an integer so that we
can set the number of nodes~$N$ to $N=N_0$.
\end{proof}

\paragraph{De Bruijn graphs}
De Bruijn graphs are used in algorithms such as
Koorde \cite{Koorde}, Distance-Halving \cite{Naor:SPAA03}, or 
ODRI \cite{Loguinov:SIGCOMM03}, and are
extensively discussed in \cite{Loguinov:SIGCOMM03,Sivarajan:TON94}.
In a de Bruijn graph, any node~$i$ is represented by an identifier string
$(i_1, \ldots, i_D)$ of $D$~symbols taken from an alphabet of size
$\Delta$. The node represented by $(i_1, \ldots, i_D)$ links to each
node represented by $(i_2, \ldots, i_{D}, x)$ for all possible values
of $x$ in the alphabet. The resulting directed graph has a fixed
out-degree $\Delta$, and a diameter $D$. 

Denote by $V'$ the set of nodes such that the identifier of each node in
$V'$ is of the form $(h,h,\ldots,h)$. Nodes in $V'$ link to themselves,
so that 
$M_i = m(\Delta - 1)$ for $i \in V'$. For nodes $i \notin V'$, the
maintenance cost $M_i$ is $M_i = m\Delta$. The next two lemmas 
will allow us to show 
that the routing cost at each node also depends on the position
of the node in the graph.

\begin{lemma}
With shortest-path routing, 
nodes~$i \in V'$ do not route any traffic, and $R_i = 0$.
\label{lemma:lowerbnd-deb-routes}
\end{lemma}
\begin{proof} 
(By contradiction.) Consider a node~$i \in V'$ with identifier
$(h,h,\ldots,h)$, and suppose $i$ routes traffic from a node~$j$ to a
node~$k$. The nodes linking to~$i$ are all the nodes with an identifier
of the form $(x,h,\ldots,h)$, for all values of $x$ in the alphabet. The
nodes linked from~$i$ are all the nodes of the form $(h,\ldots,h,y)$
for all values of $y$ in the alphabet.
Therefore, there exists $x_0$ and $y_0$ such that traffic from node~$j$
to node $k$ follows a path $\mathcal{P} = (x_0, h, \ldots, h) \rightarrow (h,
h, \ldots, h) \rightarrow (h, h, \ldots, y_0)$. Because, in a de Bruijn
graph, there is an edge between 
$(x_0, h, \ldots, h)$ and $(h, h, \ldots, y_0)$, traffic
using the path $\mathcal{P}$ between $j$ and $k$ does not follows the
shortest path. We arrive to a contradiction, which proves that 
$i$ does not route any traffic.
\end{proof}

\begin{lemma}
The number of routes $L_i$
passing through a given node $i$ is bounded by $L_i \leq L_{\max}$ with
$$
L_{\max} = \frac{(D-1)(\Delta^{D+2}-(\Delta-1)^2)-D\Delta^{D+1}+\Delta^2}{(\Delta-1)^2} \ .
$$
The bound is tight, since it can be reached 
when $\Delta \geq D$ for the node $(0, 1, 2, \ldots, D-1)$.
\label{lemma:upperbnd-deb-routes}
\end{lemma}
\begin{proof}
The proof follows the spirit of the proof used in \cite{Sivarajan:TON94}
to bound the maximum number of routes passing through a given edge. In
a de Bruijn graph, by construction, each node maps to an identifier
string of length $D$, and each path of length $k$ hops maps to a string
of length $D+k$, where each substring of $D$ consecutive symbols
corresponds to a different hop \cite{Loguinov:SIGCOMM03}. Thus,
determining an upper bound on the number of paths of length $k$ that
pass through a given node~$i$ is equivalent to computing the maximum
number, $l_k$, of strings of length $D+k$ that include node~$i$'s
identifier, $\sigma_i = (i_1, \ldots, i_{D})$, as a substring. In each 
string of length $D+k$ corresponding to a paths including $i$, where
$i$ is neither the source nor the destination of the path, the substring
$\sigma_i$ can start at one of $(k-1)$ positions $(2,\ldots, k)$. There
are $\Delta$ possible choices for each of the $k$ symbols in the string
of length $D+k$ that are not part of the substring $\sigma_i$.
As a result, 
$$
l_k \leq (k-1)\Delta^k \ .
$$
With shortest path routing, the set of all paths going through node~$i$
include all paths of length $D+k$ with $k \in [1, D]$. So,
\begin{eqnarray}
L_i & \leq & \sum_{k=1}^{k=D} l_k  \leq \sum_{k=1}^{k=D} (k-1)\Delta^k \nonumber\\
& \leq & \frac{(D-1)\Delta^{D+2}-D\Delta^{D+1}+\Delta^2}{(\Delta - 1)^2}\ . 
\label{eq:intermed-pi-bound}
\end{eqnarray}
We improve the bound given in Eqn.~(\ref{eq:intermed-pi-bound})
by considering the strings of length $2D$ that are of the form
$\sigma^*\sigma^*$, where $\sigma^*$ is a string of length~$D$. 
Strings of the form $\sigma^*\sigma^*$ 
denote a cycle $\sigma^* \rightarrow \sigma^*$, and cannot
be a shortest path in a de Bruijn graph. Hence, we can subtract the
number of the strings $\sigma^*\sigma^*$ from the bound
in Eqn.~(\ref{eq:intermed-pi-bound}). Because $\sigma_i = (i_1,
\ldots, i_{D})$ is a substring of $\sigma^*\sigma^*$ of length~$D$,
$\sigma^*$ has to be a circular permutation of $\sigma_i$, for instance
$(i_{D-1},i_{D}, i_1, \ldots, i_{D-2})$. Since $i$ does not route any
traffic when $i$ is the source of traffic, $\sigma^* \neq \sigma_i$.
Thus, there are only $(D-1)$ possibilities for $\sigma^*$, and $(D-1)$
strings $\sigma^*\sigma^*$. Subtracting $(D-1)$ from the
bound in Eqn.~(\ref{eq:intermed-pi-bound}) yields $L_{\max}$.

\end{proof}

From Lemmas~\ref{lemma:lowerbnd-deb-routes} and
\ref{lemma:upperbnd-deb-routes}, we infer that, in a de Bruijn
graph, for any $i$, $j$ and $k$, $0 \leq \Pr[\chi_{i,j}(k) = 1]
\leq L_{\max}/N^2$. Because $\chi_{i,j}(k)$ is a binary function,
$\Pr[\chi_{i,j}(k) = 1] = E[\chi_{i,j}]$, and we finally obtain $0 \leq
R_i \leq R_{\max}$ with
$$
R_{\max} = \frac{rL_{\max}}{N^2} \ . 
$$
We next compute upper and lower bounds on the access cost. To derive
a tight upper bound on $A_i$, consider a node~$i \in V'$. Node~$i$
links to itself and has only $(\Delta - 1)$ neighbors. Each neighbor of~$i$ 
has itself $\Delta$ neighbors, so that there are $\Delta(\Delta -1)$
nodes~$k$ such that $t_{i,k} = 2$. By iteration and substitution
in Eqn.~(\ref{eq:access}), we get, after simplification, $A_i \leq
A_{\max}$, with
$$
A_{\max} = a\frac{D\Delta^{D+1}-(D+1)\Delta^D+1}{N(\Delta-1)} \ ,\nonumber 
$$
and $A_i = A_{\max}$ for nodes in $V'$.

Now, consider that each node~$i$ has at most $\Delta$ neighbors.
Then, node~$i$ has at most $\Delta^2$ nodes at distance 2,
at most $\Delta^3$ nodes at distance 3, and so forth. Hence,
there are at least $\Delta^D-\sum_{k=0}^{D-1}\Delta^k$
nodes at the maximum distance of $D$ from node~$i$. We get
$$
A_i \geq \frac{a}{N}\left(\sum_{k=1}^{D-1}k\Delta^k +
D\left(\Delta^D-\sum_{k=0}^{D-1}\Delta^k\right)\right) \ , 
$$
which reduces
to $A_i \geq A_{\min}$, with
$$
A_{\min} = \frac{a}{N}\left(D\Delta^D+\frac{D}{\Delta-1}-\frac{\Delta(\Delta^{D}-1)}{(\Delta - 1)^2} \right)  . \nonumber 
$$
It can be shown that $A_i = A_{\min}$ for the node $(0, 1,
\ldots, D-1)$ when $\Delta \geq D$.

Note that, the expressions for both $A_{\min}$ and $A_{\max}$ 
can be further simplified for
$N=\Delta^D$, that is, when the identifier space is fully populated. 

\paragraph{$D$-dimensional tori}
We next consider $D$-dimensional tori, as in CAN \cite{CAN},  
where each node is represented by $D$ Cartesian coordinates,
and has $2D$ neighbors, for a maintenance cost of $M_i = 2mD$ for
any $i$. 

Routing at each node is implemented by greedy forwarding to
the neighbor with the shortest Euclidean distance to the destination.
We assume here that each node is in charge of an equal portion of the
$D$-dimensional space.
From \cite{CAN}, we know that the average length of a routing
path is $\frac{D}{4}N^{1/D}$ hops.\footnote{Loguinov et al.
\cite{Loguinov:SIGCOMM03} refined that result by distinguishing
between odd and even values of $N$.} Because we assume that the
$D$-dimensional torus is equally partitioned, we conclude by symmetry, 
that for all~$i$, 
$$ 
A_{i} = a\frac{D}{4}N^{1/D} \ . 
$$

\begin{figure}
\begin{center}
\shortstack{
\includegraphics[width=0.25\textwidth]{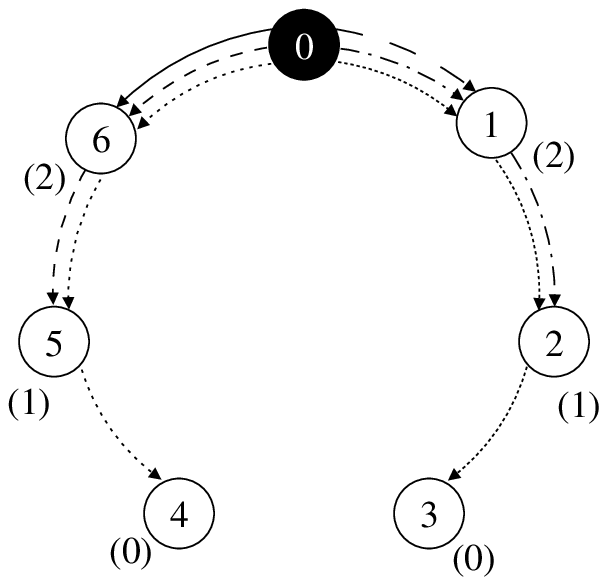}
}
\shortstack{
\includegraphics[width=0.25\textwidth]{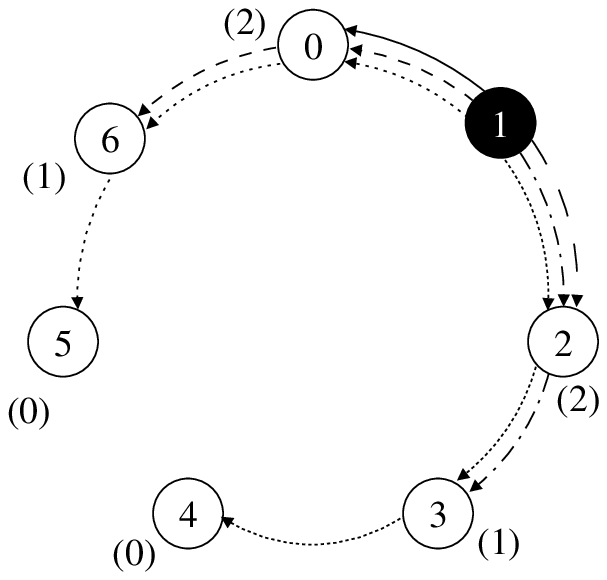}
}
\parbox[b]{0.2\textwidth}{
\centerline{$\ldots$}
\vspace{0.5in}
}
\shortstack{
\includegraphics[width=0.25\textwidth]{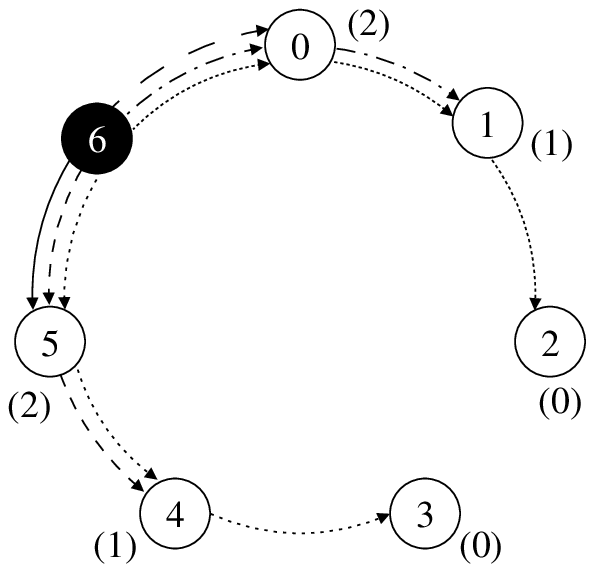}
}
\caption{\label{fig:can-proof-ring}
Routing in a ring. The numbers in parentheses represent the number of
routes originating from the black node that pass through each node.
}
\end{center}
\end{figure}
To determine the routing cost $R_i$, we compute the number of routes
passing through a given node~$i$, or {\em node loading}, as a function
$L_{i,D}$ of the dimension~$D$. With our assumption that the $D$-torus
is equally partitioned, $L_{i,D}$ is the same for all~$i$ by symmetry.
We next compute $L_{i,D}$ by induction on the dimension $D$.

\subparagraph{\bf Base case $(D=1)$.} For $D=1$, the $D$-torus is a 
ring, as depicted
in Figure~\ref{fig:can-proof-ring} for $N=7$. Each of the diagrams in
the figure corresponds to a case where the source of all requests,
represented by a black node, is held fixed. The numbers in each node
($0, \ldots, 6$) represent the node coordinate, the different line styles
represent the different routes to all destinations, and the numbers
in parentheses denote the number of routes originating from the fixed
source that pass through each of the other nodes. As shown in the figure, shifting the source of all requests from 0 to $1, \ldots, 6$ only results in shifting the number of routes that pass through each node. Hence, 
the node
loading $L_{i,1}$ at each node~$i$, is equal to the sum of the number
of routes passing through {\em each} node when the source is held fixed.
In the figure, for $N=7$, we have for any~$i$, $L_{i,1}=0+1+2+2+1+0 =
6$.
More generally, for $N$ odd, the sum
of the number of routes passing through each node is equal to
\begin{eqnarray}
L_{i,1} &=& 2\left(1+2+\ldots+\left(\frac{N-1}{2}-1\right)\right)\nonumber\\
        &=&\frac{(N-1)(N-3)}{4}\ , \label{eq:can-l1-odd} 
\end{eqnarray}
and for $N$ even, is given by 
\begin{eqnarray}
L_{i,1} &=& \left(1+2+\ldots+\left(\frac{N}{2}-1\right)\right)
+\left(1+2+\ldots+\left(\frac{N}{2}-2\right)\right)\nonumber\\
        &=&\frac{(N-2)^2}{4}\ . \label{eq:can-l1-even} 
\end{eqnarray}
We can express Eqs.~(\ref{eq:can-l1-odd}) and (\ref{eq:can-l1-even}) in
a more compact form, which holds for any $N$,
\begin{equation}
L_{i,1} = \left(\left\lfloor\frac{N}{2}\right\rfloor-1\right)
\left(\left\lceil\frac{N}{2}\right\rceil-1\right) \ .
\label{eq:can-l1}
\end{equation}

\begin{figure}
\begin{center}
\includegraphics[width=0.8\textwidth]{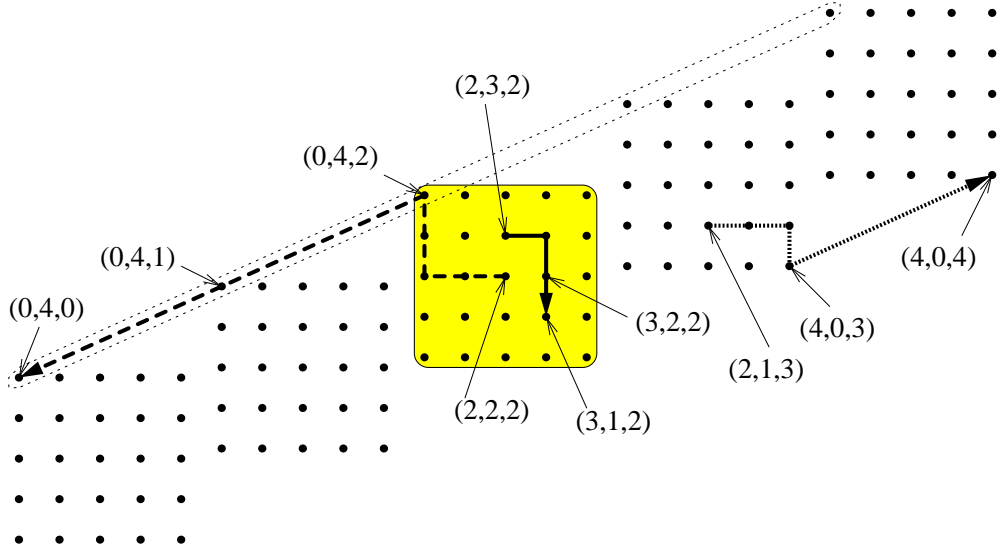}
\caption{\label{fig:can-proof-cube}
Routing in 3-torus. Coordinates are corrected one at a time, first along
the horizontal axis, then along the vertical axis, and finally along the diagonal axis.
}
\end{center}
\end{figure}
\subparagraph{\bf General case $(D>1)$.} 
The key observation to compute the number of routes $L_{i,D}$
passing through each node~$i$ for $D>1$, is that there are several
equivalent shortest paths along the Cartesian coordinates, because
the coordinates of two consecutive nodes in a path cannot 
differ in more than one dimension.
Consider for instance, for $D=2$, going from node~(0,0) to node~(1,1):
both $\mathcal{P}_1 = (0,0) \rightarrow (1,0) \rightarrow (1,1)$
and $\mathcal{P}_2 = (0,0) \rightarrow (0,1) \rightarrow (1,1)$ are
equivalent shortest paths. Therefore, we can always pick the path that corrects
coordinates successively, starting with the first coordinate, i.e.,
$\mathcal{P}_1$ in the above example.

Consider a $D$-torus, as represented for $D=3$  
in Figure~\ref{fig:can-proof-cube}, where each of the $N=125$ nodes
is represented by a dot. The figure illustrate how requests are routed by
correcting coordinates successively, with the example of three different
paths, $(2,3,2) \rightarrow (3,1,2)$, $(2,2,2) \rightarrow (0,4,0)$, 
and $(2,1,3) \rightarrow (4,0,4)$.
For any node~$k$, we compute the number of routes passing through $k$.  
We denote the source of the route as node~$i$, and the destination of   
the route as node~$j$. We have $i \neq j \neq k$. We further denote the 
coordinates of $i$, $j$, and $k$ by $(i_1, \ldots, i_D)$, $(j_1, \ldots,
j_D)$, and $(k_1, \ldots, k_D)$.
We distinguish between the only three possibilities for $k$ that are 
allowed by the routing scheme that corrects coordinates one at a time:
\begin{enumerate}
\item{Node~$k$ has the same $D$-th coordinate as both the source~$i$
and the destination~$j$, i.e., $i_D = j_D = k_D$. In other words, the
route $\mathcal{P} = i \rightarrow j$ is entirely contained within a
$(D-1)$-torus. This case is illustrated in the figure for the route
represented by a solid line going from $i = (2,3,2)$ to $j = (3,1,2)$
through $k = (3,2,2)$. The corresponding $(D-1)$-torus containing
$i$, $j$ and $k$ is denoted by the shaded box. By definition of the
node loading, the node loading resulting from all possible paths
$\mathcal{P}$ contained in a $(D-1)$-torus is equal to $L_{i,D-1}$.
There are $n$~different such $(D-1)$-tori in the $D$-torus under
consideration, one for each possible value of $i_D = j_D = k_D$. So, the
total load incurred on each node by all paths which remain contained
within a $(D-1)$-torus is equal to $nL_{i,D-1}$. }
\item{Nodes~$i$, $j$ and $k$ all differ in their $D$-th coordinate,
i.e., $i_D \neq j_D \neq k_D$. Because coordinates are corrected one at
a time, for any $l \in (1, \ldots, D-1)$, we must have $k_l = j_l$. This
case is illustrated in the figure for the route represented by a dashed
line, going from $i = (2,2,2)$ to $j = (0,4,0)$, through $k = (0,4,1)$.
Since $k_l = j_l$, nodes~$j$ and $k$ belong to the same ring where only
the $D$-th coordinate varies. Such a ring is represented in the figure
by the dotted curve. From node~$k$'s perspective, routing traffic from
$i$ to $j$ is equivalent to routing traffic between nodes~$i'$ and $j$,
where node~$i'$ satisfies $i'_l = k_l$ for any $l \in (1, \ldots, D-1)$,
and $i'_D = i_D$. (In the figure, the coordinates of node~$i'$ are
$(0,4,2)$.) From our hypothesis $k_D \neq i_D$, we have $k_D \neq i'_D$,
which implies $k \neq i'$. Therefore, computing the number of routes passing
through node~$k$ coming from $i$ 
is equivalent to computing the number of routes passing
through node~$k$ and originating from node~$i'$. Summing over all 
possible destination nodes~$j$, the computation of all routes passing through
$k$ and originating from all nodes~$i'$ in the same ring as $j$ and $k$ 
is identical to the computation of the node loading in the
base case $D=1$. So, the load imposed on node~$k$ is equal to $L_{i,1}$.
Now, summing over all possible nodes~$i$ is equivalent to summing over
all possible rings where only the $D$-th coordinate varies. There are
are $n^{D-1}$ such rings in the $D$-torus. We conclude that the total
load incurred on each node $k$ by the paths going from all $i$ to all
$j$ satisfying $i_D \neq j_D \neq k_D$ is equal to $n^{D-1}L_{i,1}$. }
\item{Node~$k$ has the same $D$-th coordinate as node~$i$, and a $D$-th
coordinate different from that of the destination~$j$. In other words, 
$i_D \neq j_D$, $i_D = k_D$. This situation is illustrated in
the figure 
for the route going from $i = (2,1,3)$ to $j = (4,0,4)$ and passing 
through $k = (4,0,3)$, and represented by a thick dotted line. In this
configuration, there are~$(n-1)$ possible choices for the destination
node~$j$ such that $j_D \neq k_D$, and $j_l = k_l$ for $l \leq D-1$.
There are $n^{D-1}-1$ possible choices for the source node~$i$ such that
$i_D = k_D$ and $i\neq j$. Hence, in this configuration, there is a
total of $(n-1)(n^{D-1}-1)$ routes passing through each node~$k$.}
\end{enumerate}
Summing the node loadings obtained in 
all three possible cases above, we obtain
$$
L_{i,D} = nL_{i,D-1}+n^{D-1}L_{i,1}+(n-1)(n^{D-1}-1) \ .
$$
Replacing $L_{i,1}$ by the expression given in Eqn.~(\ref{eq:can-l1}),
using $n=N^{1/D}$, and removing the recursion in the above relationship, we
obtain, for any node~$i$,
$$
L_{i,D} = N^{\frac{D-1}{D}}\left(D\left(N^{\frac{1}{D}}-1+\left(\left\lfloor\frac{N^\frac{1}{D}}{2}\right\rfloor-1\right)
\left(\left\lceil\frac{N^\frac{1}{D}}{2}\right\rceil-1\right)\right)-N^{\frac{1}{D}}\right)+1 \ .
$$
For all~$i$, $R_i$ immediately follows from $L_{i,D}$ with
$$
R_i = r\frac{L_{i,D}}{N^2} \ .
$$

\paragraph{Plaxton trees}
We next consider the variant of Plaxton trees \cite{Plaxton99} used in
Pastry \cite{Pastry} or Tapestry \cite{Tapestry}. Nodes are represented
by a string $(i_1, \ldots, i_D)$ of $D$~digits in base~$\Delta$. Each
node is connected to $D(\Delta-1)$ distinct neighbors of the form $(i_1,
\ldots, i_{l-1}, x, y_{l+1},\ldots,y_{D})$, for $l=1\ldots D$, and
$x\neq i_l \in \{0,\ldots,\Delta-1\}$.\footnote{For $\Delta = 2$,
this geometry reduces to a hypercube.} The resulting maintenance cost is
$M_i = m D(\Delta-1)$.

Among the different possibilities
for the remaining coordinates $y_{l+1},\ldots, y_{D}$, the protocols
generally select a node that is nearby according to a spatial proximity metric.
We here assume that the spatial distribution of the nodes is uniform,
and that the identifier space is fully populated (i.e., $N = \Delta^D$), 
which enables us to
pick $y_{l+1}=i_{l+1},\ldots, y_{D}=i_{D}$.
Thus, two nodes $i$ and $j$ at a distance of $k$
hops differ in $k$ digits. There are $D\choose k$ ways of choosing
which digits are different, and each such digit can take any of
$(\Delta-1)$ values. So, for a given node~$i$, there are ${D\choose
k}(\Delta-1)^k$ nodes that are at distance~$k$ from~$i$. Multiplying
by the total number of nodes $N = \Delta^D$, and dividing by the total
number of paths $N^2$, we infer that, for all $i$, $j$, and $k$,
we have 
\begin{equation}
\Pr[t_{i,j}=k] = \frac{{D\choose k}(\Delta - 1)^k}{N} \ .
\label{eq:tij-lpm}
\end{equation} 
Now, for any $i$ and
$j$ such that $t_{i,j} = l$, because routes are unique, there are
exactly $(l-1)$~different nodes on the path between $i$ and $j$. So, the
probability that a node~$k$ picked at random is on the path from $i$
to $j$ is 
\begin{equation}
\Pr[\chi_{i,j}(k) = 1 | t_{i,j} = l] = \frac{l-1}{N} \ .
\label{eq:chi-knowing}
\end{equation}
The total probability theorem tells us that 
$$
\Pr[\chi_{i,j}(k) = 1] = 
\sum_{l=1}^{D}\Pr[\chi_{i,j}(k) = 1 | t_{i,j} = l]\cdot\Pr[t_{i,j}=l] \ .
$$
Substituting with the expressions obtained for $\Pr[t_{i,j}=l]$ and $\Pr[\chi_{i,j}(k) = 1 | t_{i,j} = l]$ in Eqs.~(\ref{eq:tij-lpm}) and
(\ref{eq:chi-knowing}) gives:
\begin{equation}
\Pr[\chi_{i,j}(k) = 1] 
 =  \frac{1}{N^2}\sum_{l=1}^{D} (l-1){D\choose l}(\Delta - 1)^l \ ,
\label{eq:plaxton-x0}
\end{equation}
which can be simplified as follows. We write:
$$
\sum_{l=1}^{D} {D\choose l}(l-1)(\Delta - 1)^l 
= (\Delta-1)^2\sum_{l=1}^{D}(l-1){D\choose l}(\Delta-1)^{l-2} \ , 
$$
and rewrite the right-hand term as a function of the derivative of a series,
$$
\sum_{l=1}^{D} {D\choose l}(l-1)(\Delta - 1)^l 
 =  (\Delta-1)^2\frac{\partial}{\partial \Delta}\left(\sum_{l=1}^{D}{D\choose l}(\Delta-1)^{l-1}\right) \ ,$$
or, equivalently,
$$
\sum_{l=1}^{D} {D\choose l}(l-1)(\Delta - 1)^l 
= (\Delta-1)^2\frac{\partial}{\partial \Delta}\left(\frac{1}{\Delta-1}\sum_{l=1}^{D}{D\choose l}1^{D-l}(\Delta-1)^{l}\right) \ .
$$
The binomial theorem allows us to simplify the above to:
$$
\sum_{l=1}^{D} {D\choose l}(l-1)(\Delta - 1)^l 
= (\Delta-1)^2\frac{\partial}{\partial \Delta}\left(\frac{1}{\Delta-1}(1+\Delta-1)^D - 1\right) \ ,
$$
which, making the partial derivative explicit, becomes, 
$$
\sum_{l=1}^{D} {D\choose l}(l-1)(\Delta - 1)^l 
 = (\Delta-1)^2\left(\frac{D\Delta^{D-1}(\Delta -1)-\Delta^D+1}{(\Delta-1)^2}\right)\nonumber \ ,
$$
and reduces to
$$
\sum_{l=1}^{D} {D\choose l}(l-1)(\Delta - 1)^l  
= \Delta^{D-1}(D(\Delta -1)-\Delta)+1 \nonumber\ .
$$
Substituting in Eqn.~(\ref{eq:plaxton-x0}) gives:
$$
\Pr[\chi_{i,j}(k) = 1] = 
\frac{\Delta^{D-1}(D(\Delta -1)-\Delta)+1}{N^2} \ , 
$$
which we multiply by $r$ to obtain
\begin{equation}
R_i = r\frac{\Delta^{D-1}(D(\Delta -1)-\Delta)+1}{N^2} \ . 
\label{eq:routing-lpm} 
\end{equation}

To compute the access cost $A_i$, we use the relationship 
$A_i = a E[t_{i,j}]$. 
We have
$$
E[t_{i,j}] = \sum_{k=1}^{D} k \Pr[t_{i,j} = k] \ ,
$$
which, using the expression for $\Pr[t_{i,j} = k]$ given in 
Eqn.~(\ref{eq:tij-lpm}), implies 
$$
E[t_{i,j}] 
= \sum_{k=0}^{D} k \frac{{D\choose k}(\Delta - 1)^k}{N} \ ,
$$
and can be expressed in terms of the derivative of a classical series:
$$
E[t_{i,j}] 
=\frac{\Delta-1}{N}\frac{\partial}{\partial\Delta}\left(\sum_{k=0}^{D} {D\choose k}(\Delta - 1)^k\right)\ .
$$
Using the binomial theorem, the series on the right-hand side collapses to
$\Delta^D$, which yields 
$$
E[t_{i,j}] 
=\frac{\Delta-1}{N}\frac{\partial(\Delta^D)}{\partial\Delta}  \ .
$$
We compute the partial derivative, and obtain
$$
E[t_{i,j}] 
=\frac{D\Delta^{D-1}(\Delta-1)}{N} \ .\nonumber
$$
Multiplying by $a$ to obtain $A_i$, we eventually get, for all~$i$,
$$
A_i = a \frac{D\Delta^{D-1}(\Delta-1)}{N} \ , 
$$
which can be simplified, using $N = \Delta^D$: 
\begin{equation}
A_i = aD \frac{\Delta-1}{\Delta} \ . 
\label{eq:access-lpm} 
\end{equation}

\paragraph{Chord rings} 
In a Chord ring \cite{Chord:TON}, nodes are represented using
a binary string (i.e., $\Delta = 2$). When the ring is fully
populated, each node~$i$ is connected to a set of $D$~neighbors,
with identifiers $((i+2^{m}) \mod 2^D)$ for $m = 0\ldots D-1$.
An analysis identical to the above yields $R_i$ and $A_i$ as in
Eqs.~(\ref{eq:routing-lpm}) and (\ref{eq:access-lpm}) for $\Delta = 2$.
Note that Eqn.~(\ref{eq:access-lpm}) with $\Delta = 2$ is confirmed by
experimental measurements \cite{Chord:TON}.
\subsection{Numerical results}
\begin{table}
\begin{center}
\begin{tabular}{|c|c|c|c|c|c|c|}\hline
$(\Delta, D)$ & $A_{\min}$ & $A_{\max}$ & $\frac{A_{\max}}{A_{\min}}$ & $R'_{\min}$ & $R_{\max}$ & $\frac{R_{\max}}{R'_{\min}}$\\\hline
(2, 9) & 7.18 & 8.00 & 1.11 & 3.89 & 17.53 & 4.51\\\hline
(3, 6) & 5.26 & 5.50 & 1.04 & 2.05 & 9.05 & 4.41\\\hline
(4, 4) & 3.56 & 3.67 & 1.03 & 5.11 & 13.87 & 2.71\\\hline
(5, 4) & 3.69 & 3.75 & 1.02 & 1.98 & 5.50 & 2.78\\\hline
(6, 3) & 2.76 & 2.80 & 1.01 & 5.38 & 9.99 & 1.86\\\hline
\end{tabular}
\caption{
\label{tab:debruijn}
Asymmetry in costs in a de Bruijn graph $(a=1, r=1000)$}
\end{center}
\end{table}

We illustrate our analysis with a few numerical results. 
In Table~\ref{tab:debruijn}, we consider five de Bruijn graphs with
different values for $\Delta$ and $D$, and $X$ and $Y$ i.i.d. uniform
random variables. Table~\ref{tab:debruijn} shows that while the access
costs of all nodes are comparable, the ratio between $R_{\max}$ and
the second best case routing cost,\footnote{That is, the minimum value
for $R_i$ over all nodes but the $\Delta$ nodes in $V'$ for which $R_i = 0$.}
$R'_{\min}$, is in general significant. 
Thus, if $r \gg a$, there can be an incentive for the nodes with
$R_i = R_{\max}$ to defect. For instance, these nodes may leave the
network and immediately come back, hoping to be assigned a different
identifier $i'\neq i$ and incurring a lower cost. Additional mechanisms, such as
enforcing a cost of entry to the network, may be required to prevent
such defections.
\begin{figure}
\begin{center}
\shortstack{
        \includegraphics[width=0.44\textwidth]{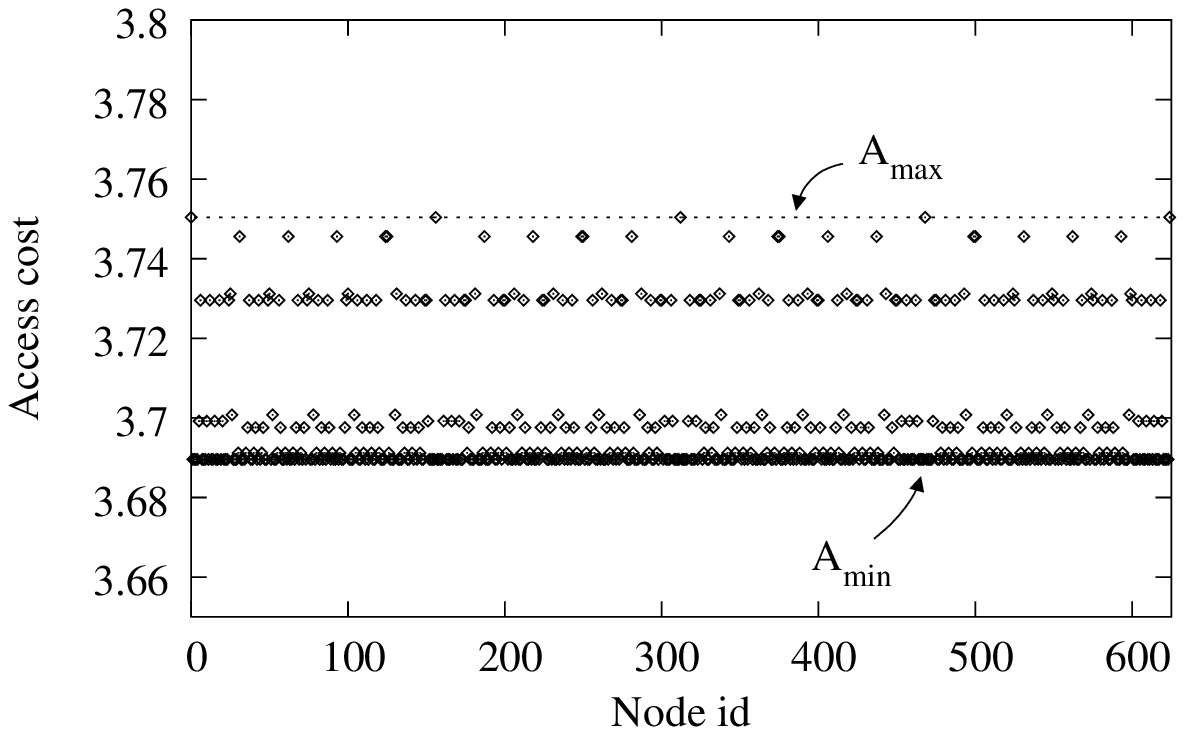}\\
        {\footnotesize (a) Access cost $(a=1)$}
}
\quad
\shortstack{
        \includegraphics[width=0.44\textwidth]{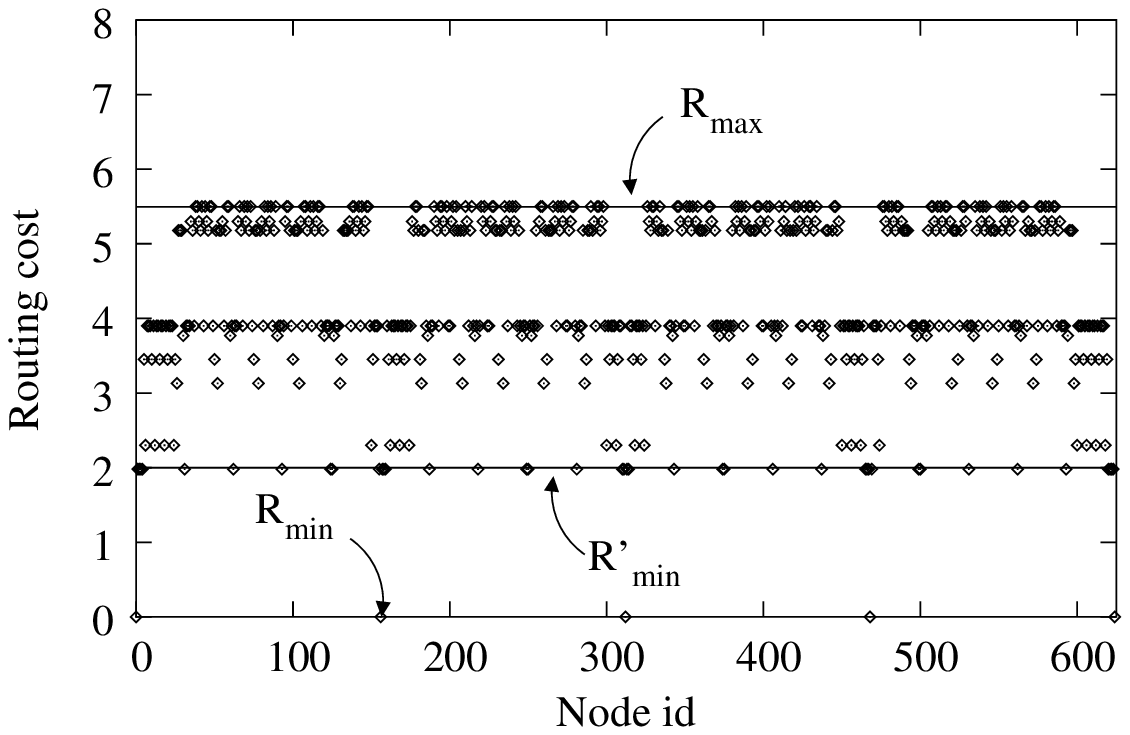}\\
        {\footnotesize (b) Routing cost $(r=1000)$}
}
\caption{
\label{fig:debruijn}
Costs in a de Bruijn network with  $\Delta = 5$, $D = 4$ and $N=625$}
\end{center}
\end{figure}
We graph the access and routing costs for the case $\Delta = 5$,
$D = 4$ and $N=625$ in Figure~\ref{fig:debruijn}. We plot the
access cost of each node in function of the node identifier in
Figure~\ref{fig:debruijn}(a), and the routing cost of each node in
function of the node identifier in Figure~\ref{fig:debruijn}(b).
Figure~\ref{fig:debruijn} further illustrates the asymmetry in costs
evidenced in Table~\ref{tab:debruijn}, by exhibiting that different
nodes have generally different access and routing costs. Therefore, in a
de Bruijn graph, there is potentially a large number of nodes that can
defect, which, in turn, may result in network instability, if defection
is characterized by leaving and immediately rejoining the network.

Next, we provide an illustration by simulation of the costs in the
different geometries. We choose $\Delta = 2$, for which the results
for Plaxton trees and Chord rings are identical. We choose $D=\{2,6\}$
for the $D$-dimensional tori, and $D=\log_{\Delta}N$ for the other
geometries. We point out that selecting a value for $D$ and $\Delta$
common to all geometries may inadvertently bias one geometry against
another. We emphasize that we only illustrate a specific example here,
without making any general comparison between different DHT geometries.

\begin{figure*}[t]
\begin{center}
\shortstack{
        \includegraphics[width=0.45\textwidth]{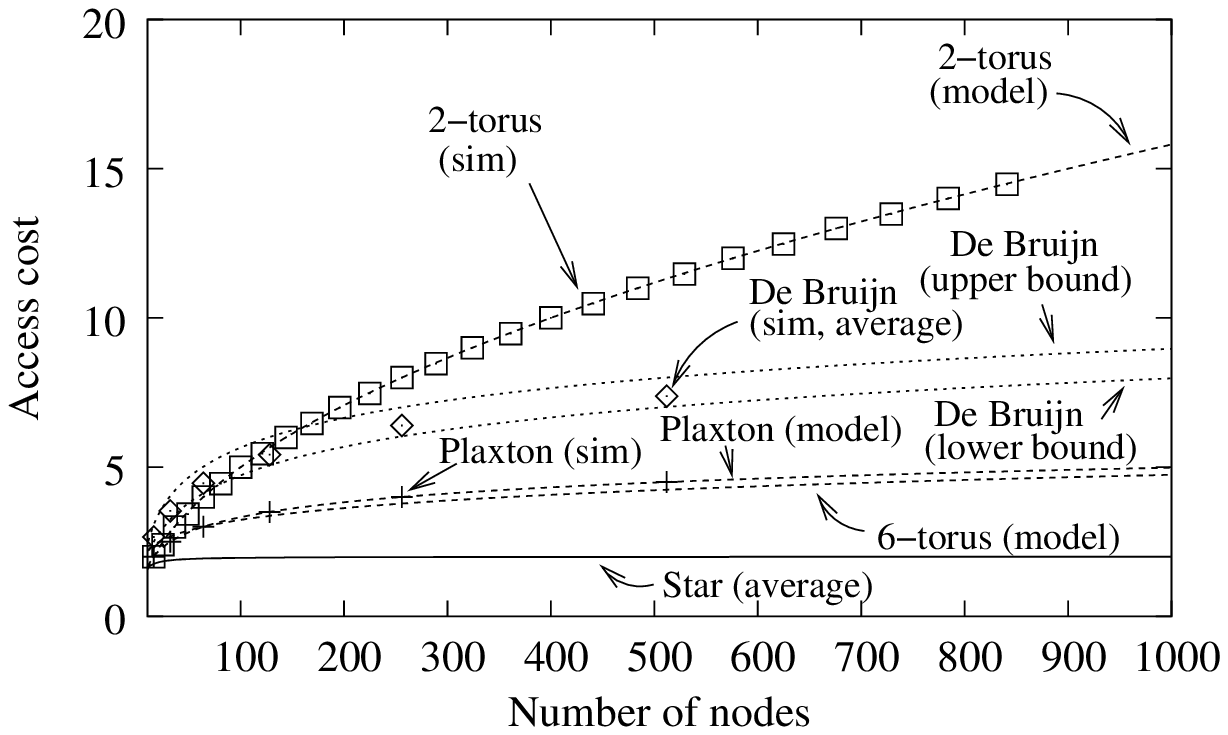}\\
        {\footnotesize (a) Access cost $(a=1)$}
}\quad
\shortstack{
        \includegraphics[width=0.45\textwidth]{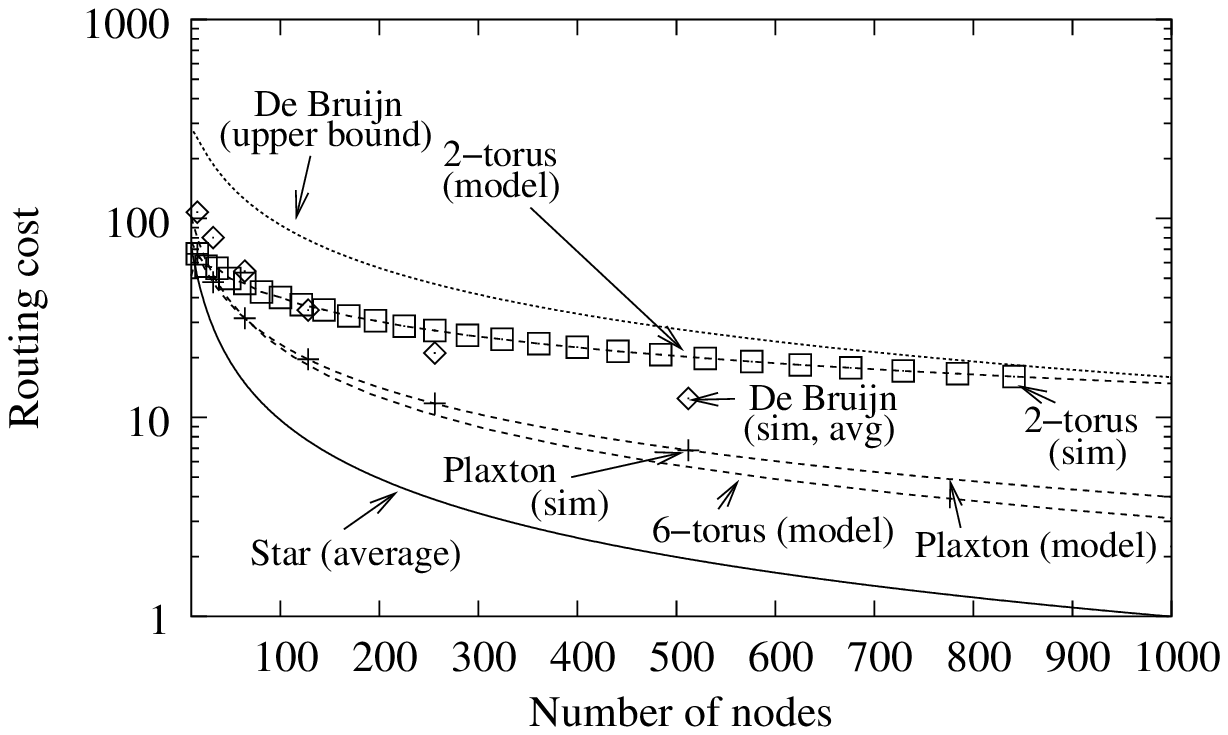}\\
        {\footnotesize (b) Routing cost $(r=1000)$}
}
\caption{
\label{fig:all}
Access and routing costs. Curves marked ``sim'' present simulation results.}
\end{center}
\end{figure*}
We vary the number of nodes between $N=10$ and $N=1000$, and, for each
value of $N$ run ten differently seeded simulations, consisting of
100,000 requests each, with $X$ and $Y$ i.i.d. uniform random variables.
We plot the access and routing costs averaged over all nodes and all
requests in Figure~\ref{fig:all}. The graphs show that our analysis is
validated by simulation, and that the star provides a lower average
cost than all the other geometries. In other words, a centralized
architecture appears more desirable to the community as a whole than
a distributed solution. However, we stress that we do not consider
robustness against attack, fault-tolerance, or potential performance
bottlenecks, all being factors that pose practical challenges in a
centralized approach, nor do we offer a mechanism creating an incentive
to be in the center of the star. While the cost model proposed here
can be used to quantify the cost incurred by adding links for a higher
resiliency to failures, we defer that study to future work.
\section{Discussion}
\label{sec:discussion}
We proposed a model, based on experienced load and node connectivity,
for the cost incurred by each peer to participate in a peer-to-peer
network. We argue such a cost model is a useful complement
to topological performance metrics \cite{Gummadi:SIGCOMM03,
Loguinov:SIGCOMM03}, in that it allows to predict disincentives to
collaborate (peers refusing to serve requests to reduce their cost),
discover possible network instabilities (peers leaving and re-joining
in hopes of lowering their cost), identify hot spots (peers with high
routing load), and characterize the efficiency of a network as a whole.

We believe however that this paper raises more questions than it
provides answers. First, we only analyzed a handful of DHT routing
geometries, and even omitted interesting geometries such as the
butterfly \cite{Viceroy}, or geometries based on the XOR metric
\cite{Kademlia}. Applying the proposed cost model to deployed
peer-to-peer systems such as Gnutella or FastTrack could yield some
insight regarding user behavior. Furthermore, for the mathematical
analysis, we used strong assumptions such as identical popularity of all
items and uniform spatial distribution of all participants. Relaxing
these assumptions is necessary to evaluate the performance of a geometry
in a realistic setting. Also, obtaining a meaningful set of values for
the parameters $(s, a, r, m)$ for a given class of applications (e.g.,
file sharing between PCs, ad-hoc routing between energy-constrained
sensor motes) remains an open problem. Finally, identifying the minimal
amount of knowledge each node should possess to devise a rational
strategy, or studying network formation with the proposed cost model are
other promising avenues for further research.

{\small
\bibliographystyle{plain}

}
\end{document}